# Is everything quantum 'spooky and weird'? An exploration of popular communication about quantum science and technology in TEDx talks

Aletta Lucia Meinsma[1,2], Sanne Willemijn Kristensen[3], W. Gudrun Reijnierse[4], Ionica Smeets[2], Julia Cramer[1,2]

[1] Leiden Institute of Physics, Faculty of Science, Leiden University, Leiden, The Netherlands
[2] Department of Science Communication & Society, Faculty of Science, Leiden University, Leiden, The Netherlands
[3] High Field Magnet Laboratory, Faculty of Science, Radboud University, Nijmegen, The Netherlands
[4] Department of Language, Literature and Communication, Vrije Universiteit Amsterdam, Amsterdam, The Netherlands

**Abstract:** Researchers point to four potential issues related to the popularisation of quantum science and technology. These include a lack of explaining underlying quantum concepts of quantum 2.0 technology, framing quantum science and technology as spooky and enigmatic, framing quantum technology narrowly in terms of public good and having a strong focus on quantum computing. To date, no research has yet assessed whether these potential issues are actually present in popular communication about quantum science. In this content analysis, we have examined the presence of these potential issues in 501 TEDx talks with quantum science and technology content. Results show that while most experts (70%) explained at least one underlying quantum concept (superposition, entanglement or contextuality) of quantum 2.0 technology, only 28% of the non-experts did so. Secondly, the spooky/enigmatic frame was present in about a quarter of the talks. Thirdly, a narrow public good frame was found, predominantly by highlighting the benefits of quantum science and technology (found in over 6 times more talks than risks). Finally, the main focus was on quantum computing at the expense of other quantum technologies. In conclusion, the proposed frames are indeed found in TEDx talks, there is indeed a focus on quantum computing, but at least experts explain underlying quantum concepts often.

## 1 Introduction

In the 20th century, the first quantum revolution took off. Scientists started to understand and apply the laws of quantum physics, which led to ground-breaking technologies such as lasers and transistors (quantum 1.0 technology). The second quantum revolution is expected to have an even bigger impact: by actively manipulating quantum effects in systems and materials, scientists are developing quantum technologies (quantum 2.0 technology) that might impact society at large (European Quantum Flagship, 2020; Stichting Quantumdelta NL, 2020; Wehner et al., 2018). Most of these quantum 2.0 technology based devices have not been realised commercially, but are developed in a research setting. This makes it hard to predict the exact impact it will have on society, although some benefits and risks have been envisioned. One of the envisioned benefits is that quantum computers can realise the development of new medicines and consequently improve healthcare (Stichting Quantumdelta NL, 2020). Also, quantum networks might impact data security by allowing for fundamentally secure communication between any two points on earth (Wehner et al., 2018). Quantum sensors, furthermore, could be crucial during military missions for navigational purposes in case GPS cannot be used in hostile environments or underground (Stichting Quantumdelta NL, 2020).

Besides potential benefits, these new quantum technologies could also pose risks for our society. For example, governments could lose their grip on criminal organisations that communicate via quantum networks. Another risk could be that large companies that develop quantum technologies might acquire even more power. Also, quantum technology might increase the power gap between developed and developing countries, when developed countries quickly adopt quantum technology while developing countries lag behind (Vermaas et al., 2019).

To maximise the societal benefits of quantum technology while minimising the risks it may pose, societal engagement could be key. One of the reasons for this is that societal engagement might lead to more socially robust solutions as a result of gaining a wider view on the impact of the technology on different social settings. Several studies, however, expect problems due to potential issues in popularising quantum science and technology (Grinbaum, 2017; Roberson et al., 2021; Roberson, 2021; Vermaas, 2017). Grinbaum (2017), for example, expects some weaknesses in the way in which popular accounts currently explain quantum 2.0 technology. He argues this might negatively influence the public's trust in quantum technologies.

The goal of our research is to identify whether potential issues in the popularisation of quantum science and technology are present in the context of TEDx talks. In these talks, speakers share their research and ideas in order to spark conversations within local communities. Furthermore, we aim to compare the talks given by quantum experts to the talks given by non-experts for containing potential issues. The next section presents our theoretical framework that includes the relevant literature on which our study is based.

## 2 Theory

*2.1 Engaging societal actors in an early stage of an emergent technology's development*

There are important reasons for engaging societal actors in early stages of an emergent technology's development. Arguments include that; 1) engaging societal actors could allow for more social contexts to shape an emergent technology (Roberson, 2021), 2) engagement could lead to more public support and less public resistance (Kurath & Gisler, 2009; Roberson et al., 2021), and 3) would fit a democratic point of view (van Dam et al., 2020). This section briefly highlights each of these arguments.

First of all, engaging societal actors could allow for more social contexts to shape an emergent technology as well as give a broader overview of the technology's impact on different social settings. Upon designing an emergent technology, the scientists, who operate in social contexts themselves, mainly imagine the use and impact of the technology they are building. Their social contexts therefore have a great influence. If different societal actors would engage in an emergent technology's development, more social contexts could shape these imaginations (Roberson, 2021).

Secondly, involving society can lead to more support and less public resistance. The history of science communication of emergent technologies reveals that some of the previous emergent technologies have had to deal with public resistance (Kurath & Gisler, 2009) and did not always lead to public benefit (Roberson et al., 2021). For instance, Roberson et al. (2021) describe an example of a new technology in agriculture, developed in the 1950s in the United States, that caused more than eighty percent of tomato-growing businesses to go into bankruptcy within five years of the technology's adoption. Businesses without adequate amounts of land were not able to benefit from the new technology. This prompted a public debate with civil society organisations about whose needs and desires the researchers (un)met.

The third argument is normative in nature. From a democratic point of view, citizens should be allowed to express their opinions and concerns on developments that largely impact their lives. Since science and technology developments can have a great impact on citizen's lives, they should be allowed to participate in the decision-making process (van Dam et al., 2020).

*2.2 Science popularisation in TEDx talks*

For societal actors to engage in early stages of an emergent technology's development, they should be able to acquire the knowledge necessary to join in (Vermaas, 2017). One platform that is reaching a worldwide audience with a wide range of topics is the TEDx platform (TED, n.d.-a.).

TEDx is a platform that hosts events similar to TED (Technology, Entertainment and Design), in which speakers give popularising talks on a variety of topics. A difference with TED is that TEDx events are locally organised. The platform's purpose is to spark conversations within local communities across the world. The local organisers invite speakers with different backgrounds and expertise to present their research and ideas in under 18 minutes, which allows the speakers to talk directly to their audience without a mediator in between (TED, n.d.-b.).

The TEDx platform targets two main audiences. The primary participants are the local communities that attend the event live and are interested in the topics discussed. The (mainly non-expert) web users who can access the talks for free via the TEDx channel on YouTube are the secondary participants (Mattiello, 2017). On May 17, 2022 the TEDx channel on YouTube had a total of 34.8 million subscribers.

These aspects make the TEDx platform an interesting context for researching the popularisation of science and technology.

*2.3 Potential issues in popularising quantum science and technology*

Our study analyses which approaches TEDx speakers use and how they frame quantum science and technology in their popular communication. Framing refers to "select[ing] some aspects of a perceived reality and mak[ing] them more salient in a communicating context, in such a way to promote a particular problem definition, causal interpretation, moral evaluation, and/or treatment recommendation" (Entman, 1993, p. 52). Highlighting particular parts while neglecting other parts of some information can have a big effect on how the audience perceives and understands this information (Entman, 1993).

We base our research on four potential issues that we found in literature. These are: 1) the use of a pragmatic approach when explaining quantum 2.0 technology (Grinbaum, 2017); 2) framing quantum science and technology as spooky and enigmatic (Vermaas, 2017); 3) the use of a narrow public good frame in relation to quantum technology (Roberson et al., 2021); and 4) the strong focus on quantum computing (Roberson, 2021).

### 2.3.1 Using a pragmatic approach for explaining quantum 2.0 technology

Grinbaum (2017) presents a first potential issue in the popularisation of quantum science and technology: he states that popular and semi-popular accounts of quantum science and technology currently use a 'pragmatic approach' when explaining quantum 2.0 technology. Such an approach skips over the explanation of underlying quantum concepts (like superposition, entanglement and contextuality; Jaeger, 2019; Nielsen & Chuang, 2010[1]), but instead focuses on protocols and how these are building blocks for quantum 2.0 technology.

While such a pragmatic approach is an efficient short-cut, there are some drawbacks associated with it as well. For example, it does not allow the public to experience the beauty of quantum theory (something that quantum physicists experience when working with the mathematics of it), while Grinbaum (2017) argues this experience is necessary for the public to gain trust in quantum technology. Grinbaum (2017) proposes that narratives could be a solution and states that *"constructing a narrative that conveys scientific content as well as provoking a feeling of beauty is the hard problem in the relations between science and society"* (p. 304). To our knowledge, no research has been performed yet to establish whether the pragmatic approach for explaining quantum 2.0 technology is indeed dominant in popular accounts.

### 2.3.2 Spooky and enigmatic frame

A second potential issue in popular communication about quantum science and technology is that the current framing of quantum science is a spooky and enigmatic one (Vermaas, 2017). An example of the spooky and enigmatic frame is a phrase coined by Einstein, who referred to the quantum mechanical concept of entanglement as 'spooky action at a distance' (translated from German *spukhafte Fernwirkung*) in one of his letters to Max Born (Einstein et al., 1971).

Vermaas (2017) argues the spooky and enigmatic frame could potentially hinder societal actors to join in on a societal debate. The reason for this is that this frame might scare societal actors and prevent them to acquire the knowledge necessary to join in on a societal debate around quantum technologies. Vermaas (2017) supports moving away from enigmatic metaphors like Schrödinger's cat (i.e. a hypothetical cat that is in a combined state of being dead and alive to depict the concept of superposition) and from saying that quantum theory is counterintuitive in nature. Instead, he proposes to introduce it as a novel information theory and to present the technological effects it might have.

In 2017, a public dialogue took place over a 3-month period with 77 participants (recruited to reflect the UK population) to find out their perceptions of quantum science and technology. Interestingly, the spooky and enigmatic frame did not seem to be caught on: none of the participants mentioned quantum science or technology to be spooky or weird (Busby et al., 2017). The question arises whether the spooky and enigmatic frame is indeed widely present in popular communication about quantum science and technology.

### 2.3.3 Narrow vs wider public good frame, benefit vs risk frame

A rhetorical analysis of the national quantum strategies of the US, the UK and Canada points out a third potential issue where a dominant frame about 'winning the quantum race' and realising economic development (the economic development/competitiveness frame) was found (Roberson et al., 2021). The Canadian national quantum strategy, for example, emphasises that nations are "racing to develop technologies that can deliver incredible capabilities which will far exceed those of conventional technologies" (Roberson et al., 2021, p. 5). Roberson et al. (2021) argue that such a frame is narrow in the sense that the uses and implications of quantum technology in and on society are not reflected upon more widely.

Roberson et al. (2021) encourage quantum physicists to use a wider 'public good' frame, i.e. a wider reflection of how quantum research could benefit and harm society. Examples of a wider public good frame have been found in the rhetoric of quantum physicists during the conference *Project Q* in 2019 that focused on political implications of quantum technology. Quantum physicists have said to be mainly "driven by goals of improving the society we live in" and are "trying to solve problems in health, energy, [and] climate change" (Roberson, 2021, p. 109), thereby framing their motivation in terms of social progress.

Both the economic development/competitiveness frame as well as the social progress frame are frames that have consistently recurred in science and technology debates (Nisbet, 2008), for example in Artificial Intelligence (AI). Cave & ÓhÉigeartaigh (2018) have assessed the risks of using the competitiveness frame (i.e. 'winning the AI race') and argue the frame is not beneficial for an inclusive, multi-stakeholder discussion on how AI can provide societal benefits while minimising risks. The social progress frame is presented as one of the alternatives that could counteract the risks the competitiveness frame poses (Cave & ÓhÉigeartaigh, 2018).

Besides focusing on the benefits for society, a wider public good frame also includes a broader reflection on the risks. Risks and benefits might influence the perception of societal actors on quantum technologies, such as Cobb (2005) found in the case of nanotechnology. Based on a random digit dialled survey of 1,536 adults, Cobb (2005) concluded that, although the framing effects are small, risk frames are slightly more influential than benefit frames. Whereas the risk frame caused respondents to acquire more negative feelings about nanotechnology, the benefit frame resulted in respondents being less angry about it. When both frames were present, the framing effects got eliminated.

We aim to determine which public good frame (i.e. narrow or wider) is more dominant in TEDx talks by comparing the use of the economic development/competitiveness versus the social progress frame, as well as the use of the benefit versus the risk frame.

*2.3.4 A focus on quantum computing*

To realise a wider public good frame, Roberson et al. (2021) suggest to focus on a wider range of quantum technology applications. This is in contrast to only focusing on the quantum computing applications of national security risks (e.g. decrypting public key cryptography) and realising economic development (e.g. big data in finance, to help calculate investment risks). In semi-structured interviews with four quantum physicists, who played important roles in designing visions for the national strategies of their countries, a focus on quantum computing got mentioned as well. One of them said that "our field is broader than quantum computing: quantum tech is much broader" (Roberson et al., 2021, p. 5). In our research, we aim to quantitatively assess whether there is a focus on quantum computing in TEDx talks.

*2.3.5 Differences in the use of frames between science experts and non-experts*

Experts and non-experts might differ in how they frame and approach their communication about emergent technologies towards a general audience. In a study on cyberinfrastructure for big data, a specific emergent information technology, Droog et al. (2020) analysed the metaphorical framing of that technology by experts and non-expert journalists. They analysed the use of metaphors in 15 US news texts (journalists) and 147 interviews with experts, and found profound differences between both groups. An example of a difference is that most experts use precise metaphorical frames (e.g., a precise metaphorical frame that contains the word 'tool' specifies the type of tool, e.g. a hammer or Swiss army knives), while the frames used by most journalists tend to be unprecise (e.g., an unprecise metaphorical frame that contains the word 'tool' does not specify which tool is referred to).

To examine whether similar differences between quantum experts and non-experts occur in popular scientific communication about quantum science and technology, we examine how often they use the four potential issues presented in sections 2.3.1 - 2.3.4.

*2.4 Aim of the study and research questions*

Despite worldwide investments into quantum science and technology (Department of Industry, Science and Resources, 2021; European Commission, 2018; Smith, 2018), there is still a lack of science communication literature on this emergent technology. That is, to our knowledge, no research to date has been performed to identify whether the potential issues of 1) using a pragmatic approach when explaining quantum 2.0 technology, 2) framing quantum science and technology as spooky and enigmatic, 3) making use of a narrow public good frame and 4) focusing on quantum computing are actually present in popular communication. To address this, we investigated whether these four potential issues are present in TEDx talks.

Our main research question was:

How do TEDx speakers frame and approach their popular communication about quantum science and technology?

To answer our main research question, we investigated the following sub-questions:

1. When mentioning quantum 2.0 technology, how often do speakers explain underlying quantum concepts on which the technology is based?
2. How often do speakers frame quantum science and technology in terms of being spooky and enigmatic?
3. How often do speakers frame quantum science and technology in terms of economic development/competitiveness and social progress?
4. How often do speakers frame quantum science and technology in terms of benefits and risks?
5. Which quantum technologies do speakers mention most often?
6. How do quantum experts and non-quantum experts compare in how they frame and approach their popular communication about quantum science and technology?

# 3 Methods

We performed a content analysis on TEDx talks that cover quantum science or technology.

*3.1 Data selection*

Figure 1 shows the sequence of our data collection. On May 18 2021, we searched for videos containing the keyword 'quantum' in the TEDx channel on YouTube by making use of the YouTube Data Application Programming Interface (API) v3 (Google Developers, n.d.). Since this API returns at most 500 videos per search, we performed individual searches per year. With 12 individual searches between 2009 (the year TEDx launched) and 2021, we acquired a total number of 4,427 TEDx talks.

We downloaded all the transcripts of the collected TEDx talks and checked whether each transcript contained our keyword 'quantum'. Whenever the keyword could not be found, we deleted the transcript. This resulted in a total of 1,002 transcripts.

Some of the talks had a transcribed or reviewed translated transcript which we refer to as 'manually transcribed'. Other transcripts had either been automatically generated by YouTube and therefore lacked punctuation, or had no English transcription available. We discarded those transcripts without an English transcription, leaving 796 automatically and 184 manually transcribed talks.

In the final step, we manually checked the transcripts for relevance to our study. When the transcript included enough substantive content about quantum science or technology, we included it in our final dataset. For example, when 'quantum' was used in 'quantum leap' or 'quantum jump' (to denote a sudden change or step forward) but with no reference to quantum physics or quantum technology, we deleted the transcript from our dataset. Table A1 in the Appendix gives an overview of the reasons for discarding transcripts. A second coder went through 10% of the data to check for relevance of the transcripts to our study, which resulted in an acceptable agreement ($\alpha$ = 0.79, 89.8%) between the two coders (Krippendorff, 2004).

As a result of the final step, we retained 531 talks, of which the 423 automatically generated transcriptions from YouTube lacked punctuation. We improved the readability of the automatically transcribed talks by automatically restoring the punctuation (Tilk & Alumäe, 2016). During the coding phases, we excluded another 7 transcripts for having a bad transcript and/or audio quality and one transcript for having appeared in the pilot already. Our final dataset thus consisted of 501 transcripts.

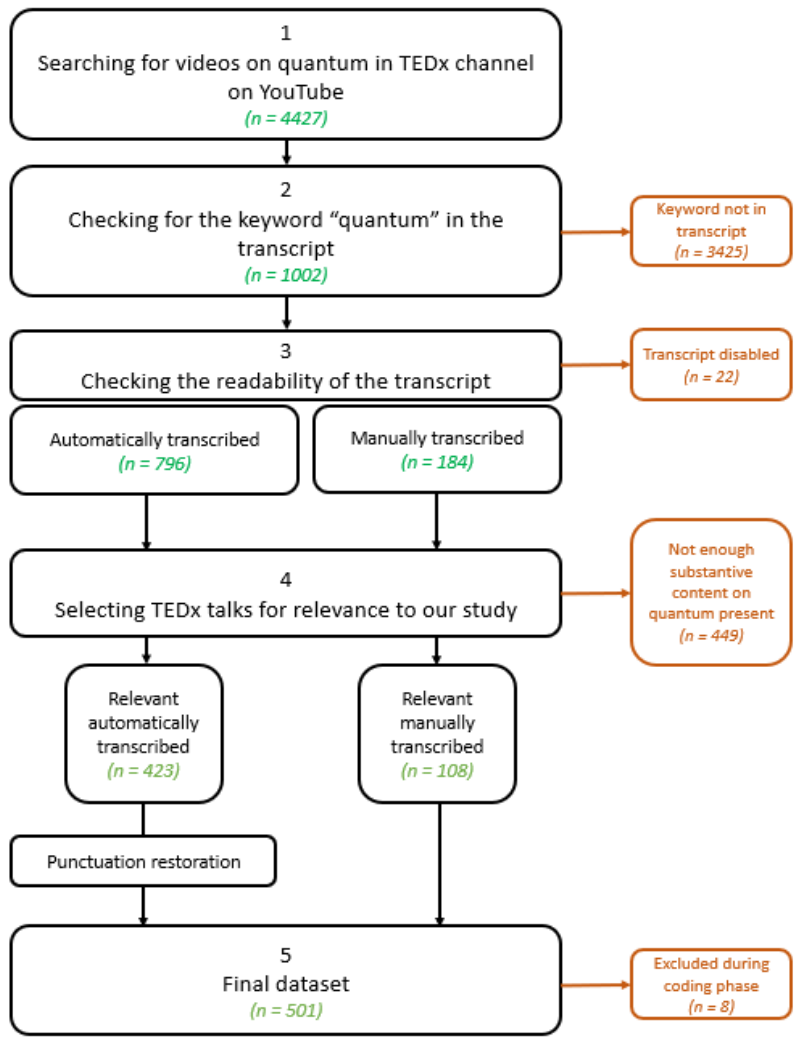

*Figure 1. Data collection method*

### 3.2 Coding

The coding consisted of two phases based on predefined codebooks. Phase 1 focussed on obtaining the descriptive data of the TEDx talk and identifying the quantum expertise and profession of the speaker at the time of the talk, for which the YouTube descriptions were consulted. Phase 2 focussed on the content of the talks, based on the YouTube transcriptions. We first identified whether quantum science and/or technology was the talk's main focus, whether the talk included a holistic viewpoint (i.e., mystical ideas related to quantum physics, often pointing out that quantum physics tells us that everything is interconnected) and if a quantum 2.0 technology indicator (i.e., a term related to quantum technology of the second quantum revolution, for example quantum computer) was present in the transcript. Afterwards, we identified for each talk whether certain quantum science explanations, frames and applications were present. To test and improve the initial codebooks, the talks of 2021 (n = 22) and 20 TED talks with quantum science and technology content that were not part of our dataset were used for a pilot study.

*3.2.1 Intercoder reliability*

To determine the reliability of our final codebooks, two coders (the first and second authors of the paper) independently went through 15% of the final dataset (76 YouTube descriptions and transcripts). In phase 1, we found an acceptable agreement between the two coders (α= 0.78, 88.6%) for identifying speakers' quantum expertise and profession. Because reliability remained relatively low in the pilot study for phase 2, the first and second coder discussed their independent codings of the final dataset to reach agreement. Before the discussion, percent agreements between 80% and 100% were achieved, except for the codes on establishing if the TEDx talk has a 'quantum' focus (78%) and determining the use of the benefit frame (61%). Even though most codes achieved high percent agreements, Krippendorff's α remained low for some of them. We think the reason is the low number of times that those codes appeared in talks (see Table A2 in the Appendix), meaning that a slight difference in interpretation between the first and second coder is already detrimental for α. By further refining the codes and with the discussion in mind, the first coder went through the remaining 85% of the transcripts.

The complete coding scheme can be found in the Appendix.

# 4 Results

*4.1 Descriptive data*

The 501 TEDx talks in our dataset were performed at TEDx events all over the world with a total of 55 different countries present in our dataset. Most talks were given in North America (n = 222 talks, 44%) and Europe (n = 173 talks, 35%) (see figure A1 in the Appendix). The talks last between 3:49 minutes and 33:19 minutes and include the word 'quantum' 8 times on average (the median is 3). Almost half (n = 236, 47%) of the 501 TEDx talks have quantum science and/or technology as their main focus. Furthermore, 73 talks (15%) include a holistic or mystical viewpoint. For example, one of the speakers mentions that quantum physics is related to mental telepathy (TEDx Talks, 2016a, 12:11):

> *"[...] over the years I've been exposed to things like quantum physics, something known as the Morphogenic field, the field of our mind, expanding far beyond our brains, a kind of mental telepathy that says there could be a tipping point that if just one more person on the planet picks up a peaceful practice, and that person could be here in this room today, that there would be an instantaneous shift in everyone's awareness and the peace from our individual hearts would be communicated directly from mind to mind to mind spontaneously to everyone, everywhere on the planet."*

Additionally, we identified the quantum expertise of the speakers at the time they presented the talk. In total, 23 speakers gave 2 or more TEDx talks, and expertise was coded by talk. Generally, the talks were given by one speaker (n = 490, 98%), but 11 talks (2%) were given by two speakers (expertise was coded for each speaker). As a result, we identified the expertise of 512 speakers. The majority of speakers were non-experts (n = 237, 46%), followed by quantum experts (n = 192, 38%). There were also relatively many speakers whose expertise we were not able to identify ('unknown', n = 83, 16%). Figure 2 shows the professions of the speakers and their related quantum expertise category in more detail. Most quantum experts (n = 153, 30%) worked at a university, institute or (inter)national research organisation. Non-quantum experts had a main expertise domain outside quantum science, for example an academic working in a different research field than quantum science. Perhaps somewhat unexpectedly, one of the non-quantum experts was even a shepherd from Pashmina, India.

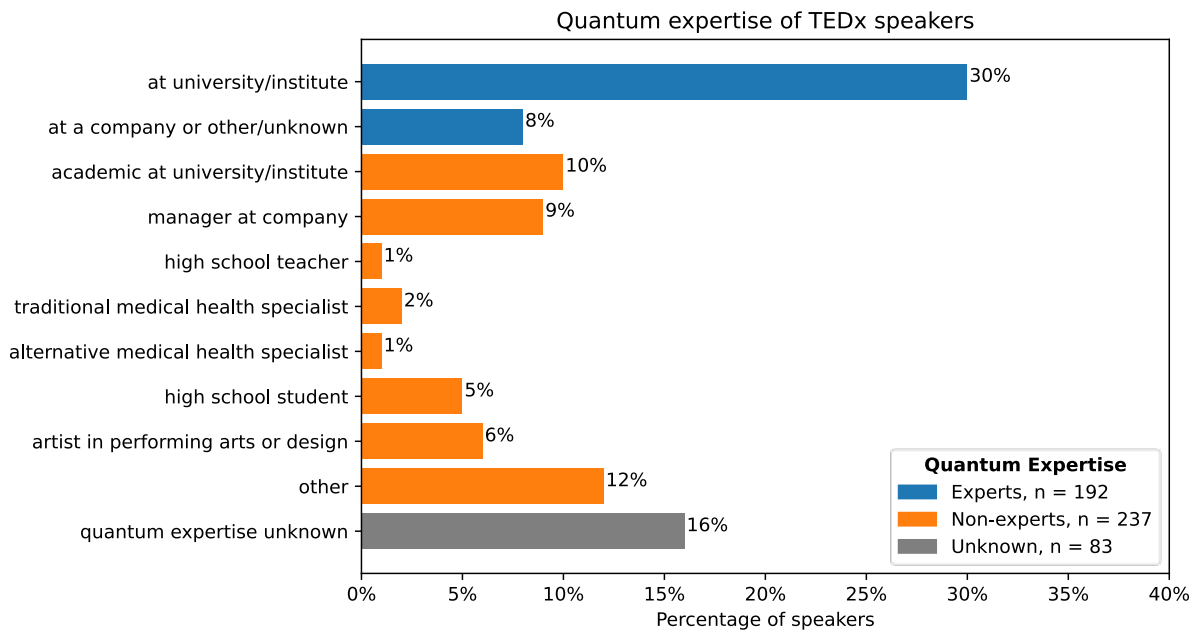

*Figure 2. The percentage of TEDx speakers per profession and related quantum expertise*

*4.2 Explanation of underlying quantum concepts when mentioning quantum 2.0 technology*

We established whether TEDx speakers made use a pragmatic approach when explaining quantum 2.0 technology, i.e. applications from the second quantum revolution such as quantum computers and quantum networks. Quantum 2.0 technology got mentioned in n = 127 talks. If those talks contained an explanation of an underlying quantum concept, i.e. superposition, entanglement or contextuality, the speaker consequently did not make use of a pragmatic approach. In more than half of the quantum 2.0 technology talks the speaker gave an explanation of at least one underlying quantum concept (n = 69, 54% of the quantum 2.0 talks).

Out of the three quantum concepts that we studied, superposition was explained most often (n = 57, 45% of the quantum 2.0 talks) followed by entanglement (n = 32, 25% of the quantum 2.0 talks) and contextuality (n = 31, 24% of the quantum 2.0 talks). An example of a superposition explanation is *"If I make this ball quantum, […], my quantum ball can be red or it could be blue or it can be red and blue at the same time. […] It's a little weird, but we call it the superposition."* (TEDx Talks, 2019, 5:10) Secondly, an example of an entanglement explanation is *"What happens if I take two of these quantum balls and I put them in a special kind of superposition state that I'm going to call an entangled state. Well, this leads to some very strong correlations between the two balls, so strong, in fact, that it no longer makes sense to talk about them as separate objects"* (TEDx Talks, 2019, 5:54).

Thirdly, an example of a contextuality explanation is *"But what happens if I try to look at this quantum ball? Well, it turns out that I, as a classical observer, cannot actually view the superposition, but very actively trying to look at the ball forces it to be either red or blue."* (TEDx Talks, 2019, 5:37).

*4.3 The spooky and enigmatic frame*

Furthermore, in our dataset, the spooky and enigmatic frame is apparent but not dominant: 115 talks (23%) framed quantum science (applications) as spooky or enigmatic or a synonym of spooky or enigmatic. Examples are *"Everything 'quantum' is spooky and weird"* (TEDx Talks, 2017, 8:06) or *"Einstein, he called this spooky action at a distance"* (TEDx Talks, 2012, 3:06).

*4.4 The economic development/competitiveness vs social progress frame, and benefits vs risks frame*

Besides the spooky and enigmatic frame, we analysed four more frames: the economic development/competitiveness frame, the social progress frame, the benefit frame, and the risk frame. Although holistic talks (n = 73, 15%) were found to frame quantum science and technology in terms of benefits, social progress and risks, they were unrelated to our research question. We therefore excluded the holistic talks from this part of the analysis. Consequently, the percentages presented in this section are with respect to the therefore relevant dataset (n = 428 talks). Table 1 gives an overview.

**Table 1**
*Frequency table of the talks that include the economic development/competitiveness frame, social progress frame, benefit frame, and risk frame*

| Frame | Total number of talks | Percentage of relevant dataset |
|---|---|---|
| Economic development / competitiveness | 23 | 5% |
| Social progress | 31 | 7% |
| Benefit | 146 | 34% |
| Risk | 22 | 5% |

*Note.* Multiple frames can appear in a talk. The percentages provided are with respect to the hereby relevant dataset (total dataset with the holistic talks excluded, n = 428 talks).

In our dataset, the amount of talks that mention the economic development/competitiveness frame (n = 23, 5%) was slightly lower to that of the talks with the social progress frame (n = 31, 7%). Of those, three talks included both frames simultaneously (1%). Examples of the economic development/competitiveness frame include *"quantum mechanics based products and services represent about more than one fifth of our gross national product"* (TEDx Talks, 2009, 15:20) and *"there is a massive race toward building a new technology called quantum computing"* (TEDx Talks, 2015, 3:19). An example of the social progress frame is *"I'm going to tell you about how to make the world a better place with quantum mechanics"* (TEDx Talks, 2016, 00:09).

A difference is more apparent between the benefit and risk frames: over 6 times more talks included the benefit frame (n = 146, 34%) compared to the risk frame (n = 22, 5%). A balanced view was provided (i.e. both frames present simultaneously) in 15 talks (4%). Benefits were most often mentioned in the life sciences & health field (n = 52, 12%), followed by energy & climate (n = 46, 11%). The security & privacy field was mentioned most often (n = 16, 4%) when a speaker highlighted a risk. In the Appendix, figure A2 shows an overview of the percentage of talks with the benefit and risk frame in specific fields.

*4.5 Technology applications of quantum science*

In total dataset, 197 talks (39%) mentioned at least one technology application of quantum science. Quantum computers and simulators were the most often mentioned technology applications of quantum science (n = 120, 24%), followed by quantum networks and cryptography (n = 30, 6%), and classical computers (n = 25, 5%). An overview is presented in Figure 3, where quantum 1.0 refers to applications from the first quantum revolution, such as lasers and smartphones, and quantum 2.0 refers to applications from the second quantum revolution (quantum 2.0 technology). The other/unknown category (n = 82, 16%) contains technologies with a link to quantum science that we identified qualitatively. The top 3 applications that we identified in this category are transistors (n = 18, 4%), the scanning tunnelling microscope (n = 10, 2%) and solar cells (n = 6, 1%).

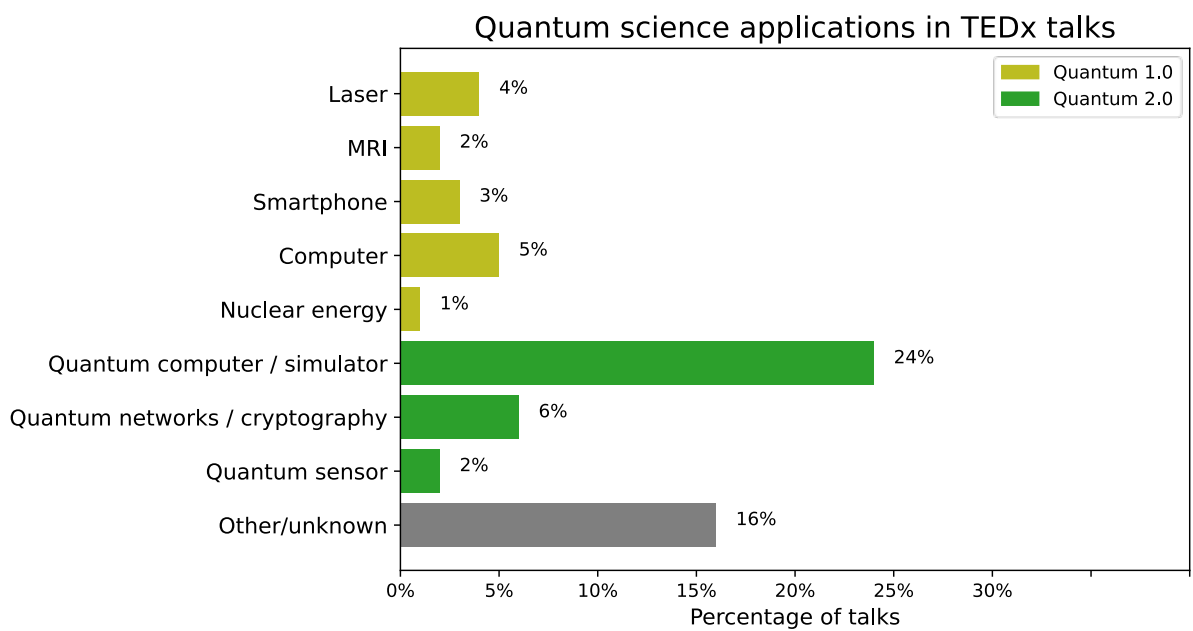

*Figure 3. The percentage of quantum science applications mentioned in TEDx talks*

*4.6 Comparison between quantum experts and non-experts*

To establish whether there are any differences between the quantum expert (n = 192) and non-expert (n = 237) groups in how they frame and approach the popularisation of quantum science and technology, we did a comparison. The group for which we were not able to determine the speaker's expertise was excluded in this comparison. The differences between the groups are tested for significance with a chi-square test. Note that this analysis only gives an exploratory view since there is a dependence between the data points.

First of all, there is a difference between the expert and non-expert group in explaining underlying quantum concepts when mentioning quantum 2.0 technology. Although 70% of the experts (n = 52 out of 72) that mention quantum 2.0 technology provided an explanation of at least one of the quantum concepts that we researched, only 28% of the non-experts (n = 13 out of 47) that mention quantum 2.0 technology did so. This difference between the groups is statistically significant, as confirmed by a chi-square test ($\chi^2(1) = 20.992, p = < 0.001$). Figure A3 in the Appendix provides a comparison between the groups per quantum concept explanation (superposition, entanglement and contextuality).

Secondly, experts (n = 58, 30% of the experts) framed quantum as spooky or enigmatic more often than non-experts (n = 39, 16% of the non-experts) ($\chi^2(1) = 11.456, p = < 0.001$).

Apart from the spooky frame, we furthermore compare the expert and non-expert use of the economic development/competitiveness, social progress, benefit and risk frame in relation to quantum science and technology. In this comparison, we again excluded the speakers that expressed holistic views in their talk (4 experts and 56 non-experts) resulting in a total of n = 188 experts and n = 181 non-experts to compare. Figure 4 shows that experts used all these frames more often than non-experts did[2].

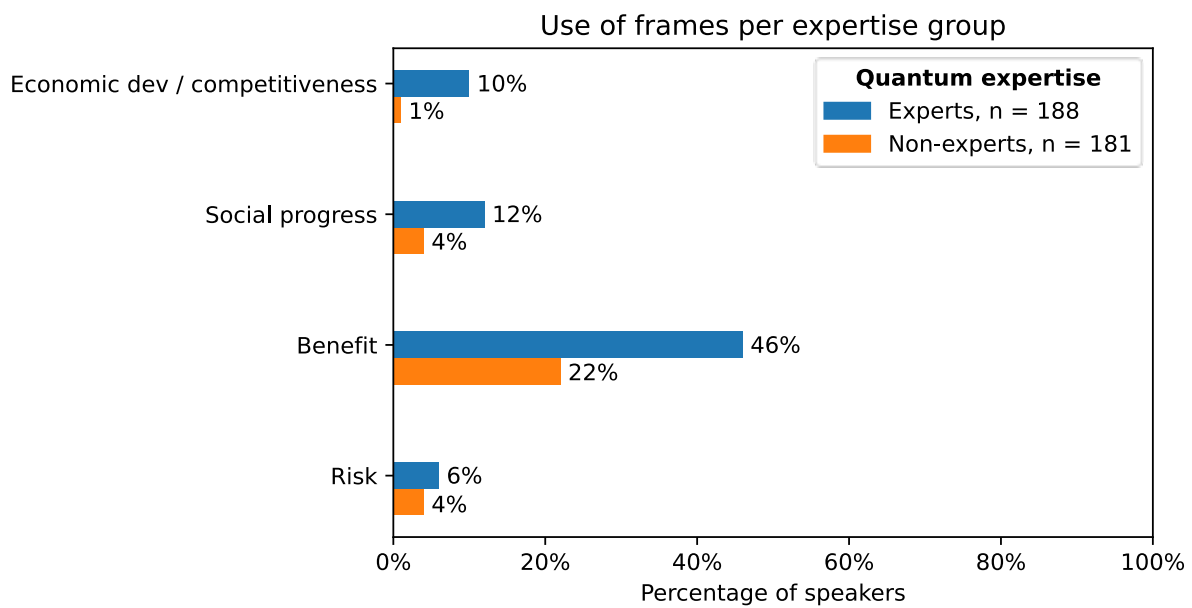

*Figure 4. The percentage of speakers per expertise group that use the economic development/competitiveness frame, the social progress frame, the benefit frame and the risk frame. Speakers that expressed holistic views in their talks are excluded from this comparison*

Finally, the quantum computer/simulator was the most often mentioned quantum science application for both experts (n = 71, 37% of their talks) and non-experts (n = 43, 18% of their talks). For both groups, there is a big gap to the second most often mentioned application (n = 24, 13% of the experts and n = 5, 2% of the non-experts mentioned quantum networks/cryptography).

# 5 Conclusion and Discussion

In this study we explored how TEDx speakers frame and approach their popular communication about quantum science and technology. To the best of our knowledge, this is the first study that quantitatively assesses whether four potential issues in popularising quantum science and technology are present in a type of communication towards a broader audience.

We performed a content analysis on 501 TEDx talks that cover quantum science or technology. Within this sample, the majority of the talks were given by non-experts, followed by quantum experts. Some experts and non-experts expressed holistic and mystic ideas on quantum science and technology. In over half of the talks that mentioned quantum 2.0 technology the speaker gave an explanation of superposition, entanglement, and/or contextuality. Experts explained underlying quantum concepts more often than non-experts did. About a quarter of the talks framed quantum as spooky or enigmatic. A narrow public good frame was apparent in the sense that risks were hardly mentioned, while benefits were communicated about much more often. Although not appearing often, the social progress frame (a wider public good frame) was mentioned slightly more often than the economic development/competitiveness frame (a narrow public good frame). Almost 4 out of 10 talks mentioned at least one quantum science application, of which quantum computers and simulators were mentioned the most often.

## 5.1 Explanations of underlying quantum concepts mostly given by experts

In our research, we investigated four possible issues in the popularisation of quantum science and technology. One of the issues might be the focus on protocols and applications of quantum 2.0 technology without explaining underlying quantum concepts, while Grinbaum (2017) argues that experiencing the beauty of quantum theory is necessary for the public to gain trust in quantum technology. We found that most experts (n = 52, 70%) did explain at least one underlying quantum concept (superposition, entanglement or contextuality). However, only 28% of the non-experts (n = 13) did so.

## 5.2 Spooky and enigmatic frame apparent but not dominant

Secondly, Vermaas (2017) argues that quantum science is being framed as spooky and enigmatic, which might hinder societal actors to participate in a societal debate about quantum technology. We found that the spooky and enigmatic frame is apparent but not dominant: 115 talks (23%) framed quantum science (applications) as spooky or enigmatic or a synonym of spooky or enigmatic. More experts than non-experts made use of this frame. There might be a relation with Albert Einstein's well-known framing of quantum entanglement as 'spooky action at a distance', although only 18 out of 115 talks specifically used this quote.

## 5.3 Beneficial framing of quantum science and technology

Benefits and risks might influence the perception of societal actors on quantum technologies, such as Cobb (2005) found. In our research, benefits of quantum technology were mentioned in over 6 times more talks than risks. Only 4% of the talks provided a balanced view, including both the benefit and risk frame which Cobb (2005) showed eliminates the framing effects in the case of nanotechnology. Almost half of the experts (n = 87, 46%) and almost a quarter of the non-experts (n = 39, 22%) framed quantum science and technology in terms of benefits, while only 6% of the experts (n = 12) and 4% of the non-experts mentioned risks. Consequently, we found a narrow public good frame, while Roberson et al. (2021) encourage quantum physicists to use a wider public good frame, i.e. a wider reflection of how quantum research could benefit and harm society. Although not appearing often, a wider public good frame was found in the sense that the social progress frame (a wider public good frame) was used slightly more often (7%) than the economic development/competitiveness frame (a narrow public good frame) (5%).

*5.4 A focus on quantum computing*

When discussing quantum science applications, TEDx speakers (both experts and non-experts) focused on quantum computing. This is in line with the complaint by one of the quantum physicists interviewed by Roberson et al. (2021) who said that: "our field is broader than quantum computing: quantum tech is much broader" (p. 5)*.* To realise a wider public good frame, speakers can give attention to a wider range of quantum 2.0 technology applications. It is expected that for example first generation quantum cryptography and specific quantum sensors have a high technology readiness level compared to other quantum technology applications (European Quantum Flagship, 2020) and therefore might have societal implications before quantum computing.

*5.5 Recommendations for further research*

With our findings on how TEDx speakers frame and approach their popular communication about quantum science and technologies, further questions on the popularisation and societal engagement of quantum technology come up. For example, journalists and experts might frame quantum technology differently from one another, similar to what Droog et al. (2020) found for cybersecurity. A content analysis on media platforms could be interesting to compare with our study on TEDx talks. Furthermore, while we mapped the presence of four potential issues of framing and approaches, their effect on the public perception is largely unknown. Does a pragmatic approach indeed have some drawbacks for quantum technology as Grinbaum (2017) suggested? Will a spooky frame (Vermaas, 2017) and a 'winning the quantum race' frame (Cave & ÓhÉigeartaigh, 2018; Roberson et al., 2021) hinder an effective societal engagement on quantum technology? Future research on the impact of such frames on public perceptions will give insight into the benefit or harm of such frames. Finally, we already encourage quantum experts to reflect on these findings, especially in giving a more balanced view on the impact of quantum technology when talking to a broader audience, to enhance an open societal discussion on the future impact of quantum technology.

*Notes*
*1. The explanations of these quantum concepts can be found in Appendix A1.*
*2. We performed a chi-square test to test for significance for the data in figure 5 (with mutually exclusive categories), but these test results are less reliable due to multiple cells having expected counts less than five.*

## Acknowledgements

We thank Sanne J.W. Willems for her valuable insights into using a chi-square test of independence and Anne Drou Roget who worked on a similar topic during her Master thesis. We acknowledge funding from the Dutch Research Council (NWO) through a Spinoza grant awarded to R. Hanson (project number SPI 63-264). This work was supported by the Dutch National Growth Fund (NGF), as part of the Quantum Delta NL programme.

# Appendix

*A1. Explanations of quantum concepts superposition, entanglement and contextuality*

Three quantum physics concepts that underlie quantum technology are superposition, entanglement and contextuality. The first concept, superposition, refers to the fact that a particle in a superposition state can be in a linear combination of states. For example, an electron in a superposition state can exist in spin states up and down at the same time. Secondly, entanglement means that two particles share a strong connection with each other - measuring one of the particles instantly affects the state of the other, even when the particles are separated by a large distance. The third concept, contextuality, denotes that performing a measurement on a quantum state affects the state irreversibly. For more information on these concepts, we would like to refer to Nielsen & Chuang (2010) and Jaeger (2019).

*A2. Reasons for deleting transcripts in step 4 of the data collection*

**Table A1**
*Reasons for deleting transcripts in step 4 of the data collection*

| Reasons for deleting a transcript in step 4 | Example | Automatically transcribed | Manually transcribed |
| --- | --- | --- | --- |
| 1. Used the keyword to metaphorically denote a sudden change or step forward | "quantum leap" "that was a quantum step up from zero" | 69 | 12 |
| 2. Mentioned the keyword in a list of other types of sciences, technologies or other terms without any further mention of it | "But if I make a list – ok, chemtrails is a bit more extreme – zodiac signs, let's see, tarot cards, quantum-psycho…" | 101 | 28 |
| 3. One or multiple persons or institutes are involved in or know about quantum science, but there is no mention of the keyword any further | "Quantum physicist Max Planck has said, 'When you change the way you look at things, the things you look at change.'" | 48 | 15 |
| 4. The speaker clarifies that s/he is not using/talking/going to talk about quantum science | "I just finished my PhD in September on the quantum photophysics of organic solar cells. That's not what my talk is on, so don't worry." "I don't mean that in the quantum mechanical sense of the term of parallel universe" | 24 | 4 |
| 5. Mentioned the keyword to indicate something's superiority / significance or a quantity of something | "One zettabyte is equal in to all the video titles on netflix x 470 million times. This is the quantum of data." | 18 | 1 |
| 6. Using keyword to indicate that a topic is very difficult | "Then stop spending that much, save a little and buy yourself that new rifle. It's not quantum physics!" | 6 | 2 |
| 7. Due to another reason, which we indicated in our annotations | E.g. making a side joke: "You, as a non-quantum observer, (Laughter), can see this band..." | 108 | 14 |

*A3. Low Krippendorff's α in Phase 2 for codes with low prevalence*

In Phase 2 of the coding scheme, Krippendorff's α remained low for some of the codes while they achieved high percent agreements. Table A2 shows the number of times those codes appeared in talks (which we think might have been detrimental for α) together with the corresponding percentage agreement and α.

**Table A2**
*The intercoder reliability for codes with a low prevalence in Phase 2*

| Code in Phase 2 | Percent agreement | α < .667 | Number of times coded in 15% of data after discussion (76 transcripts) |
|---|---|---|---|
| Laser | 93% | 0.63 | 4 |
| Smartphone | 84% | 0.25 | 3 |
| Computer | 80% | 0.40 | 6 |
| Quantum sensor | 96% | -0.13 | 1 |
| Social progress | 82% | 0.382 | 6 |
| Economic development / competitiveness | 91% | 0.319 | 4 |
| Risk frame | 91% | 0.414 | 4 |
| Contextuality explanation | 92% | 0.657 | 12 |

*A4. The complete coding scheme*

## 1. Identification

*1. Who is coding?*
1 = Coder A, 2 = Coder B

*2. What is the video ID of the TEDx talk?*

*3. What is the title of the TEDx talk?*
If no title given, then insert 0.

*4. In what year was the TEDx talk published on the TEDx YouTube Channel?*

*5. What is the name of the TEDx event?*

*6. Is the transcription automatically generated?*
0 = no, 1 = yes

*7. Is the TEDx talk flagged in the TEDx talk description? This means that it falls outside TEDx's curatorial guidelines.*
0 = no, 1 = yes

## 2. Speaker identification

*8. What is the number of speakers in the TEDx talk?*
If there are 2 speakers, code 9 and 10 for each speaker separately.

*9. What is the name of the speaker?*

*10. What is the quantum expertise and current profession of the speaker as provided in the YouTube description?*

Quantum experts are scientists (undergraduates and graduates excluded) and leaders (e.g. a founder, director, CEO, chairman, chief, etc.) at a university, institute, research initiative, start-up, or another organization working in or having worked in the field of quantum nanotechnology or another field in which quantum science plays a role. These include:

- Examples of fields in quantum nanotechnology: quantum technology, quantum information processing, quantum computational processing, nanotechnology
- Other examples of fields in which quantum science plays a role: quantum mechanics, quantum field theory, string theory, quantum optics, quantum cosmology (incl Big Bang theory and the black hole information paradox), quantum gravity, particle physics, high energy physics, photonics, condensed matter physics, nuclear physics, post-quantum cryptography

*If someone is described as a quantum expert: code 1 or 2.*
*If the quantum expertise is undefined, code 10.*
*If someone has no quantum expertise and multiple current professions apply: choose the first profession in the list or if other: code 11.*

1 = quantum expert currently working at a university/research institute/(inter)national research organisation

> Exclude (under)graduate students (=11). E.g. of an international research organisation: CERN, TNO.

2 = quantum expert currently working at a company or other/unknown

> If other/unknown: specify the other/unknown in the 'comment' column. E.g. of other/unknown: a retired quantum expert, or a quantum physicist working at the EU Quantum Flagship (=international initiative).

3 = academic / leader at a university/research institute/(inter)national research organisation, main expertise domain lies outside quantum science.

> Exclude (under)graduate students (=11), exclude leaders of spiritual institutes (=7)

4 = (executive) manager / leader at a company

> Exclude quantum experts (=2), exclude founders of research institutes (=1 or 3) or other initiatives (=2, 7 or 11), like leaders of spiritual institutes (=7)

5 = high school teacher

> Exclude university lecturers (=3)

6 = traditional medical health specialist working in health care

> E.g. a surgeon, psychologist, psychiatrist

7 = alternative medical health specialist working in a holistic medical field or in a spiritual centre

> E.g. a holistic doctor, sound therapist, founder of a spiritual centre

8 = high school student

9 = artist currently working in performing arts (e.g. musician) or in design (e.g. designer, architect)

> Include artists that bridge quantum science and arts with each another.

10 = (quantum expertise) unknown

> No information on the speaker is given or the quantum expertise is undefined. The latter means that the speaker works in a 'broad' field that includes both classical and quantum science and the specific research is not specified, e.g. 'a professor in physics' without a specification of the research s/he works on.
>
> Examples of 'broad' fields are astrophysics, physics, cosmology - incl dark energy and black holes-, biology, chemistry.
>
> When the current profession is known: include in comment section.

11 = other, namely...

> E.g. an undergraduate or graduate student, a monk, an officer
>
> Also code 11 if the profession only partly fits one of the non-quantum expert categories above, e.g. a biomedical engineer but it is unclear whether s/he works at a university or in a company.
>
> If someone has multiple professions: choose the first one you come across in the talk.

### 3. Focus of the TEDx talks

*11. Is a main topic of the TEDx talk 'quantum'?*

0 = no

1 = yes, on quantum nanotechnology

> The talk focusses on quantum science applications such as nanotechnology, quantum technology 1.0 (technology based on quantum transport) or quantum technology 2.0 (manipulating and reading out single quantum states, falls into one of the application domains: quantum computing & simulation, quantum communication, and quantum sensing & metrology).

2 = yes, on quantum science or a topic in which quantum science plays a role (exclude quantum nanotechnology)

> The talk is about pure quantum science, or a topic in which quantum science plays a role. Examples are a talk on quantum mechanics, quantum field theory, string theory, quantum optics, quantum cosmology (incl Big Bang theory and the black hole information paradox), quantum gravity, particle physics, high energy physics, photonics, condensed matter physics, nuclear physics, post-quantum cryptography.

*12. Does the TEDx talk include a holistic viewpoint?*

> An example of a holistic viewpoint is when a speaker mentions that quantum mechanics tells us that everything is interconnected, for example that an illness in one causes an illness in others too.

0 = no, 1 = yes

*13. Is a quantum technology 2.0 indicator present in the transcript?*

> Quantum technology 2.0 indicators include the term 'quantum' and belong to one of the following application domains: quantum computing & simulation, quantum communication, and/or quantum sensing & metrology.
>
> Examples: quantum technology, quantum computer, quantum algorithm.

0 = no, 1 = yes

*14. If a quantum technology 2.0 indicator is present (13 is 1=yes), quote the quantum technology 2.0 indicator. If multiple quantum technology indicators are present, only quote the first one.*

### 4. Researching quantum science applications, frames and explanations

**Quantum science applications**

*15. Is there at least one technology application of quantum science present in the transcript?*

Include: quantum technologies 1.0 (applications based on quantum physics such as the laser or smartphones), quantum technologies 2.0 (applications such as quantum computers and quantum networks), as well as other nanotechnologies (other applications in the nanometer-size range such as nanotubes).

Do not include: when the transcript includes classical technologies based on quantum physics (i.e. quantum technologies 1.0), but the link with quantum physics is absent in the transcript.

0 = no, 1 = yes

*16. Quantum science applications: which technology application(s) of quantum science is/are mentioned?*

For each technology, indicate whether it is mentioned (0 = no, 1 = yes). Codes 1 to 5 are examples of quantum technology 1.0 and codes 6 to 8 of quantum technology 2.0.

1 = Laser
2 = MRI scanner
3 = Smartphone
4 = Computer
5 = Nuclear energy
6 = Quantum computer or quantum simulator
7 = Quantum network, quantum internet, quantum cryptography
8 = Quantum sensor
9 = Other / category unsure

**Quantum science frames**

*17. Is the frame 'Quantum science (applications) is (are) spooky or enigmatic' present?*

<span style="color:green">Include:</span> when a synonym of spooky or enigmatic refers to quantum science, a quantum science principle or a quantum science application. E.g.: quantum entanglement, quantum tunneling, string theory, quantum simulator.

<span style="color:red">Do not include:</span> when a synonym of spooky or enigmatic refers to concepts that follow the rules of quantum science, like photons and atoms, or that might be related to quantum science, like dark energy.

Examples of synonyms of spooky or enigmatic are: chilling, creepy, eerie, ghostly, mysterious, ominous, scary, supernatural, uncanny, weird, strange, ambiguous, cryptic, obscure

0 = no, 1 = yes

*18. If the spooky frame is present (17 is 1=yes), quote the sentence that includes the frame. If multiple sentences apply, only quote the first one.*

*19. Is the frame 'Quantum science (applications) lead(s) to social progress' present?*
The 'social progress frame' in relation to quantum science and its applications means that quantum would mean something good for society and should be developed and deployed in such a way.

<span style="color:green">Include:</span> Quantum nanotechnologies could help solve societal problems (e.g. climate change and global malnutrition), or quantum nanotechnologies would impact society in a positive way.

<span style="color:red">Do not include:</span> If there is no mention of society being impacted, or if it is not clear whether society is impacted in a positive way (i.e. whether quantum nanotechnology brings social progress).
For example: "Quantum nanotechnologies would change the lives of many": do not include if it does not becomes clear that the lives of many will change in a positive way.

0 = no, 1 = yes

*20. If the social progress frame is present (19 is 1=yes), quote the sentence that includes the frame. If multiple sentences apply, only quote the first one.*

*21. Is the frame 'economic development/competitiveness' present?*
The 'economic development/competitiveness frame' in relation to quantum science and its applications means that various parties are in competition to develop quantum nanotechnology, there is a quantum race going on. These parties invest heavily in quantum nanotechnologies. Quantum nanotechnology will provide economic growth, and will therefore have an impact on all kinds of industries.
<u>*Note: both the social progress frame and the economic development frame can appear in a text: one frame does not exclude the other.*</u>

For example: Nations should invest in quantum nanotechnologies in order to win the quantum race

0 = no, 1 = yes

*22. If the economic development/competitiveness frame is present (21 is 1=yes), quote the sentence that includes the frame. If multiple sentences apply, only quote the first one.*

*23. Are benefits of quantum science (applications) mentioned?*
>   For example: quantum computers will be able to solve specific simulation and optimalisation problems exponentially faster than supercomputers currently can.

0 = no, 1 = yes

*24. If benefits are mentioned (23 is 1=yes), quote the sentence that includes the benefit. If multiple sentences apply, only quote the first one.*

*25. If benefits are mentioned (23 is 1=yes): are specific fields to which the benefits apply mentioned?*
For each field, indicate whether it is mentioned (0 = no, 1 = yes).

1 = Life sciences & health
>   For example: using quantum simulators to develop new medicines.

2 = Finance
>   For example: using quantum computers to run optimalisation algorithms to help model the risks of investment decisions.

3 = Logistics
>   For example: using quantum computers to run optimalisation algorithms to model the traffic flow.

4 = Security & privacy
>   For example: data will be inherently safe against eavesdropping with the use of quantum networks.

5 = Defense
>   For example: using quantum sensors during military missions for navigational purposes (in case GPS cannot be used, for example in hostile environments or underground).

6 = Energy & climate
>   For example: using quantum computers to model better batteries.

7 = Agriculture, water and food
>   For example: using quantum sensors to detect water contamination.

8 = other

*26. Are risks of quantum science (applications) mentioned?*
>   For example: quantum computers will impact the financial system, because cyber criminals can hack online banking; terrorists will be able to create new weapons by using a quantum computer; the power difference between poor and rich countries becomes bigger once the rich countries own quantum technologies whereas the poor do not.

*Note: both the risk and benefit frame can appear in a text: one frame does not exclude the other.*
0= no, 1 = yes

*27. If risks are mentioned (26 is 1=yes), quote the sentence that includes the risk. If multiple sentences apply, only quote the first one.*

*28. If risks are mentioned (26 is 1=yes): are specific fields to which the risks apply mentioned?*
For each field, indicate whether it is mentioned (0 = no, 1 = yes).
1 = Life sciences & health
>   For example: terrorists using quantum simulators for bioterrorism purposes.

2 = Finance
>   For example: cyber criminals using quantum computers to hack into online banking.

3 = Logistics
>   For example: terrorists using quantum computers to get access to air and railroad traffic controls.

4 = Security & privacy
  For example: governments losing their grip on criminal organizations that make use of quantum communication.
5 = Defense
  For example: terrorists using quantum computers to gain access to military information.
6 = Energy & climate
  For example: terrorists using quantum computers to hack into energy plants.
7 = Agriculture, water and food
  For example: terrorists using quantum computers to hack into water supplies and water management.
8 = other

**Quantum science explanations**

*29. Is the word 'superposition' mentioned?*
0 = no, 1 = yes

*30. Is an explanation provided of the quantum science principle: superposition?*
  A particle in a superposition state can be in multiple quantum states at the same time. For example, when an electron is in a superposition state, it can exist in spin states up and down at the same time.

  Include: something is 0 and 1 at the same time.     Do not include: when there is no explanation present, but the speaker just mentions the word 'superposition'. For example: a qubit can be in a superposition of 0 and 1.

0 = no, 1 = yes

*31. If an explanation of superposition is provided (29 is 1=yes), quote the sentence that includes the explanation. If multiple sentences apply, only quote the first one.*

*32. Is the word 'entanglement' mentioned?*
0 = no, 1 = yes

*33. Is an explanation provided of the quantum science principle: entanglement?*
  Two entangled particles share an extremely strong connection with each other - measuring one of the particles instantly affects the state of the other, even when the particles are separated by a large distance. In other words: entangled particles can only be described by the quantum state for the entire system, and not by their individual quantum states.

  Include: two entangled particles affect each     Do not include: holistic explanations like
  other even when they are very far apart. Also     everything is interconnected.
  include if an analogy or metaphor is provided to
  explain the principle.
0 = no, 1 = yes

*34. If an explanation of entanglement is provided (32 is 1=yes), quote the sentence that includes the explanation. If multiple sentences apply, only quote the first one.*

*35. Is an explanation provided of the quantum science principle: contextuality?*
  Contextuality means that "outcomes of measurements [depend] on other measurements on the self same system" (Jaeger, 2019, p. 2). This means that when performing a measurement on a quantum state, that measurement affects the quantum state irreversibly.

*36. If an explanation of contextuality is provided (35 is 1=yes), quote the sentence that includes the explanation. If multiple sentences apply, only quote the first one.*

*A5. The origins of the TEDx talks in our dataset*

The TEDx talks have been presented at events that were held in 55 different countries. Figure A1 shows the percentage of TEDx talks per continent and table A3 presents the top 10 countries that appear in our dataset most often.

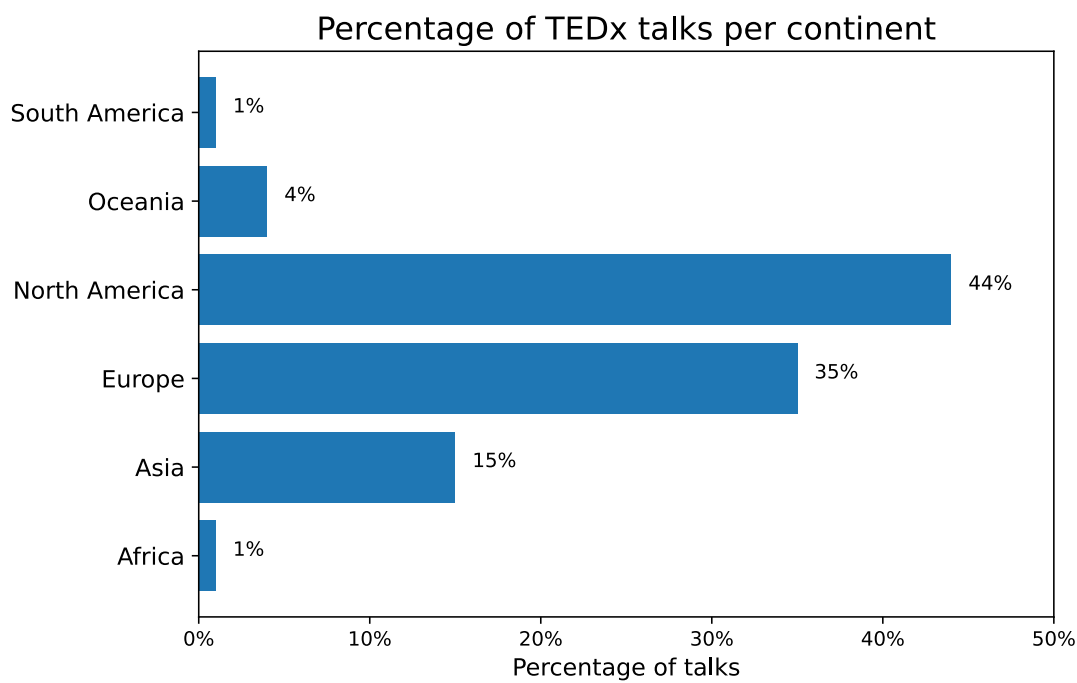

*Figure A1. The percentage of TEDx talks in our dataset for each continent*

**Table A3**
*The top 10 countries at which the TEDx talks in our dataset were presented*

| Country | Number of TEDx talks |
| --- | --- |
| United States | 181 |
| United Kingdom | 46 |
| Canada | 37 |
| India | 37 |
| Australia | 20 |
| Netherlands | 18 |
| Belgium | 14 |
| Germany | 13 |
| Italy | 11 |
| France | 9 |

*A6. Comparison between experts and non-experts*

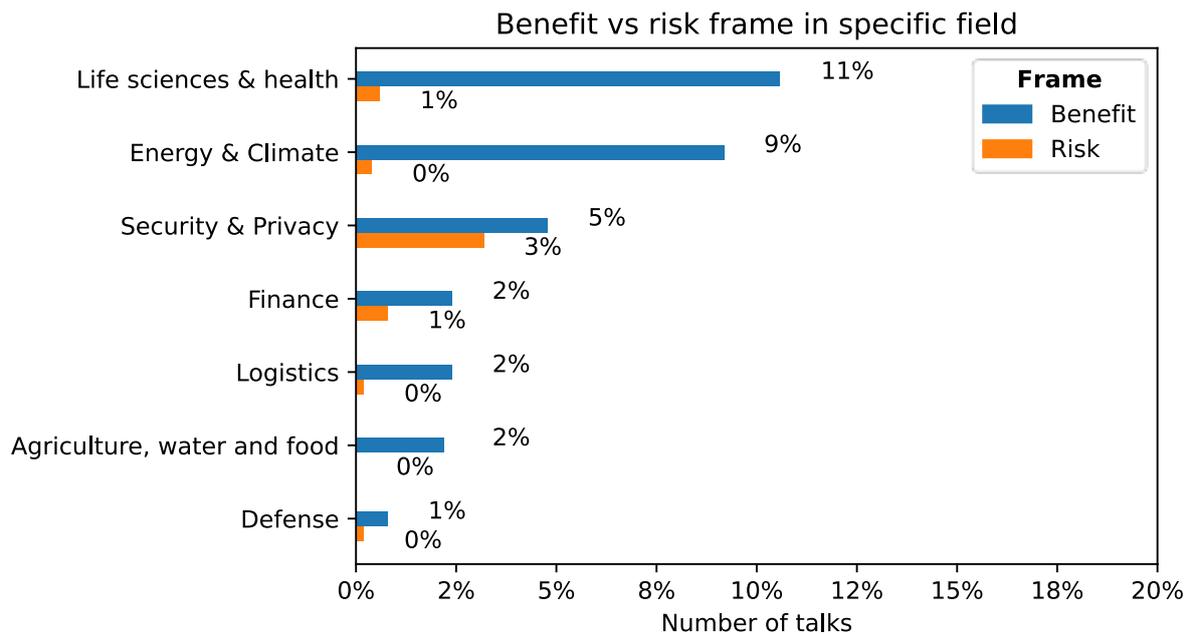

*Figure A2. The benefits and risk frame for specific fields. These fields are based on Stichting Quantum Delta NL (2020) prediction of fields that will benefit from quantum 2.0 technology*

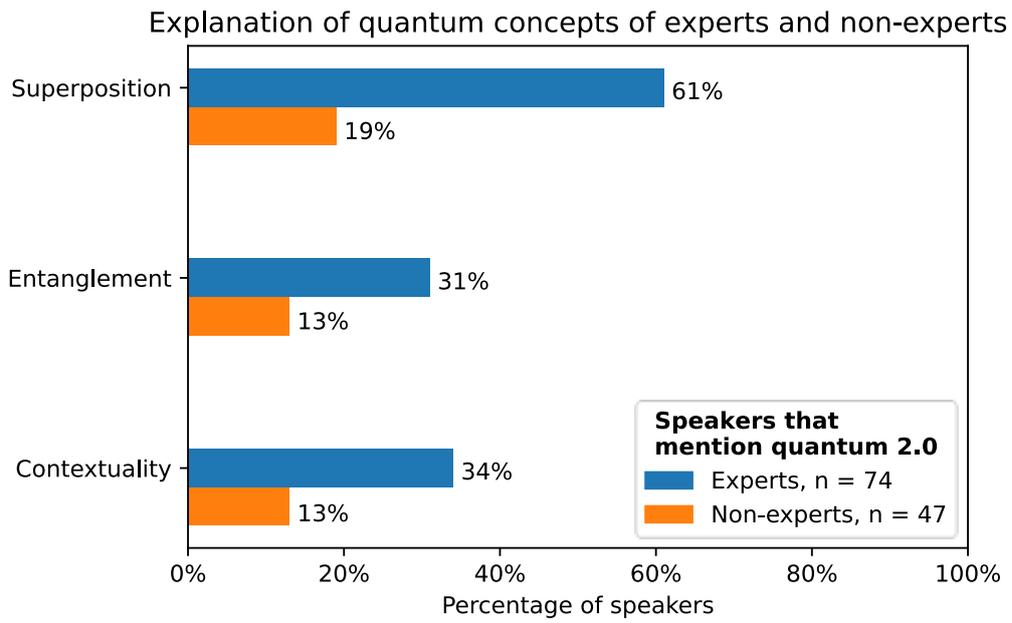

*Figure A3. The percentage of speakers per expertise group that provide an explanation of an underlying quantum concept when mentioning quantum 2.0 technology*